\documentclass[preprint,amsmath,amssymb,nofootinbib,tightenlines,floatfix,latterpaper]{revtex4}
\usepackage[dvips]{graphics}
\usepackage{latexsym}
\usepackage{slashed}
\newcommand{\beq}{\begin{equation}}
\newcommand{\eeq}{\end{equation}}
\newcommand{\beqr}{\begin{eqnarray}}
\newcommand{\eeqr}{\end{eqnarray}}
\begin{document}
\title{Nonlinearity  in Oscillatory Flows over Sand Ripple}
\author{Ruma Dutta\footnote{Electronic address: ruma@mps.ohio\-state.edu}
} 

\affiliation {Dept of Civil engineering \& Geodetic Science, The Ohio State Univ} 
 
\author{Ethan Kubatko\footnote{Electronic Address: kubatko.3@osu.edu} 
} 
\affiliation{ Dept. of Civil Engineering \& Geodetic Science, The Ohio 
State University}
\begin {abstract}
\label{sec: Abstract}

 In this report, We investigated the nonlinear phenomena in 
the study of flow dynamics of velocity component. 
In our studies, we observed nonlinear term in the 
vertical component of velocity 
by vittori et al \cite{vit} 
The time series simulation of 
vertical component which increases with Reynold stress value. 
 We developed direct numerical simulation under two dimensional grid system 
to study the flow dynamics and vorticity parameter. 
Flow pattern and flow dynamics  near wavy boundary wall in the victinity of 
ripple bottom was readdressed under direct numerical 
simualtion(DNS) framework. 
Both horizontal and vertical component of 
fluid velocity were studied under pulating force of flow.
 Vorticity is calculated under complex framework by taking into 
higher order interaction term.  
we tried to carry out similar simulation with 
same particle ejection in the viscous bed using DNS simulation for 
pulsating flow. 
Our focus was to observe particle motion using DNS 
simulation and study the 
particle phase under vortex structures formed here.  
\end{abstract}

\maketitle

\section{Introduction}
 Many observations were made cencerning a complex bed form
pattern in the victinity of offshore region. 
Temporal chaos in fluid turbulence is the symptotatic
of spatial chaos. The chaos in turbulence can be studied 
can be studied by many numerical approaches. 

As widely discussed, the chaotic pattern is detected in 
turbulence both for fixed geometric configuration increasing 
the flow Reynolds number and for fixed characteristic 
of oscillatory flow increasing the amplitude of the wall
waviness. 
From an analysis of the 
The tide driven current in the offshore region is the 
m
major source  
sediment transport in beach areas. The current deflects 
toward the crests due to increase in bottom friction 
with a decreasing water depth. Hence the cross ridge 
velocity increases to satisfy continuity whereas the 
along ridge velocity decreases owing to increase of 
bottom friction. Huthnance was the first to present a 
mathematical description using simple flow model. He used a 
simplified version of depth average shallow water 
equation with a power law relationship for sediment transport
corrected for downhill gravitational transport. 

and turbulence pattern under complex bed-fluid intearction 
mechanism. In fact, 
vast sea water(both offshore and near shore) is not clear 
Underneath sea with complex bed form having ripple is 
observed occasionally.  \cite{sleath}\\ 
Indeed many observations were made concerning a complex 
bed form pattern in the victinity of offshore region. 
The bottom topography is indeed very complex in nature. 
These ripples are formed due to oscillatory nature of 
turbulence flow over sand bed form in a complex manner 
where the bottom topography often takes form of  
The chaotic phenomena related to turbulence studies were observed by 
\cite{blond}
 brick or tile pattern depending on the complexity of 
the bed-fluid interaction under turbulence flow. 
Flow becomes more complex near the shore region 
where the flow due to 
oscillatory nature affects the ripple formation.  
It is believed that waves are mainly responsible for sediment
transport where currents
carry the entrained sediments away. Field observations focussing 
tide driven region indicate the presence of symmetrical 
and asymmetrical waves with crest almost perpendicular to
the direction of main current and characterized by 
wavelengths of few hundered metres. 
In this region, full 
understanding of
turbulence with oscillatory flow is needed to 
understand the sediment flow. Turbulent
fluctuations due to oscillatory nature of flow are usually 
confined within a thin oscillating boundary layer. 
This situation makes very difficult to take 
 experimental data 
accurately in this region which makes sometimes the
interpretation of data controversial. There are different 
numerical approaches to study these problems such as 
Reynolds Averaged Naveir Stokes Equation(RANS), 
`Direct Numerical Simulation(DNS). 
LES model is based on solving Navier Stokes equation 
with large eddy function and small eddies are neglected 
in this calculation. It well known for its simplicity and low 
computer cost but can not give more insight 
in the complex phenomena. On the other hand, Reynolds Averaged 
Navier Stokes Equation alsoo known as RANS model is based on 
On the other hand, DNS simulation does not involve any 
turbulence model 
but uses unsteady flow using grid system that are sufficiently 
fine to resolve all scales of motion.  
 A first attempt to explain the mechanism of this flow 
made by Hara an Mei \cite{hara}  who developed a 
three dimensional model 
investigating the stability of Stokes layer 
induced by sea wave. 
The associated steady state along the ripple surface showed a 
tendency to accumulate sediment particle in various pattern.  
B Blondeaux \& Vittori \cite{blon}  
tried to model sand ripple based on brick pattern form on the 
basis of three dimensional simulation model. They studied 
three dimensional vortex structure in the 
oscillatory form of flow under two dimensional ripple bed. 
In both cases, a unidirectional oscillatory flow is 
considered and fluid particles are considered 
on top of the bottom boundary to oscillate to and fro.   
Ripple at sea bed affects the sediment transport rate and 
cause additional energy dissipation 
enhancing mixing in the vicinity of ripple. However 
the detail knowledge of the flow structure and the 
dynamics of vortex structure generated by flow 
seperation is not clear in this region. 
Recently Scandura \& Blondeaux \cite {Scan}  studied by means 
of numerical simulations, the flow induced by wavy wall under uniform 
oscillatory motion. 
They observed in the simulation that velocity is periodic under weak flow
and vorticity is shed just above the crest which has a 
tendency for pitchfork 
bifurcation above critical value of velocity. \\
\indent  In our numerical simulation, we tried to 
readdress the problem 
of chaos in velocity and vorticity
field and observed similar bifurcation at critical value of velocity.  
Our direct numerical simulation is based on finite difference scheme 
for oscillatory flow of fluid and we observe  
development of bifurcation in the normal and streamline 
flow component which increases with $ u_{\rm 0} $. The nonlinearity
nature of the vertical component of the velocity also shows similar 
pattern.

\section{Formulation of the Problem}
 
The problem was formulated in the following way, 
We consider incompressible fluid of density $\rho$ and kinematic 
viscosity $\nu$ induced close to a 
wavy wall by a uniform oscillating 
pressure gradient. We define Cartesian orthogonal corordinate 
We start with  Navier Stokes equation for an 
incompressible fluid flow in rectangular $(x,y,z)$ 
coordinates. We also consider wall profile 
described parametrically by the relationship \\

\begin{equation}
   y= - {\frac {h}{2}} \lbrack cos({\it k}{\xi}) + 
    {\sum_{n=1}}^N c_{\rm n} cos({\gamma}_{\rm n})  \rbrack 
\eeq
\beq
   x= {\xi} + \frac{h}{2} \lbrack sin(k{\xi}) + 
 {\sum_{n=1}}^N c_{\rm n} sin({\gamma_n}) \rbrack
\eeq
where $ {\rm k} = \frac{2{\pi}}{l} $ is the wavenumber of the 
waviness. $\xi$ is a dummy variable and 
$ \gamma_{\rm n}  = nk{\xi} + {\phi}_{\rm n} $. 
\begin{align}
&\ \frac{\partial u}{\partial t} + {\bf u }\cdot \bigtriangledown u
 = -\frac{1}{\rho} \frac{\partial p}{\partial x}
 + {\nu}\bigtriangledown^{2}u + F^{(x)} \\
&\ \frac{\partial w}{\partial t} + {\bf u} \cdot \bigtriangledown w =
 -\frac{1}{\rho} \frac{\partial p}{\partial z} +  
    {\nu}\bigtriangledown^{2}w \\
&\ \bigtriangledown \cdot {\bf u} = 0
\end{align}

   For Direct numerical simulation algorithm, we choose 
   collocated, nonstaggered grid system. The algorithm we 
\begin{align}
-\frac{1}{\rho} \frac{\partial p}{\partial y}  + 
{\nu}\bigtriangledown^{2}v \\
&\ \frac{\partial w}{\partial t} + {\bf u} \cdot \bigtriangledown w =
 -\frac{1}{\rho} \frac{\partial p}{\partial z} +  
{\nu}\bigtriangledown^{2}v \\&\ \bigtriangledown \cdot {\bf u} = 0
\end{align}

where the field velocities (\it{ u,v,w}) are along 
(\it{x,y,z}) directions repectively.

\indent 
For the sediment particle, the basic equation is controlled by 
spherical particle moving under gravity in viscous fluid.

\subsection{Discussion and Conclusion of the Results}

\indent 
We consider the flow of an incompressible viscous fluid of 
density $\rho$ and kinematic viscosity $\nu$ induced close 
to wavy wall by a uniform oscillating pressure gradient. 
The nonlinear fluctuation and periodicity was observed here 
for velocity component both in streamline and vertical flow field.   
The nonlinearity increases with $u_{\rm o} $ above the 
threshold value. Whwn the shear stress experienced by the interface 
betwen the flowing fluid and the resting particles is low, the 
flow is unable to entrain the particles lying on the bed, which 
then remains immobile. As the shear stress increases,

\section{Acknowledgement}
This work was supported by Naval Research Lab Grant.

\end{document}